# Prepared MR Elastography

Tanguy Boucneau[1], Brice Fernandez[2], Luc Darrasse[1], and Xavier Maître[1]

[1]Université Paris-Saclay, CEA, CNRS, Inserm, BioMaps, Orsay, France
[2] Applications & Workflow, GE Healthcare, , France

## Synopsis

By decoupling motion and spatial encoding, magnitude contrast MR Elastography could be performed for the first time at ultrashort echo times (12 µs). On the basis of a motion-sensitizing magnetization preparation, the available total magnetic moment is sensitized to the motion induced in the tissues so the information can be efficiently carried over by the MR signal magnitude when the selected imaging pulse sequence is applied. The new paradigm allows also for shorter total acquisition times as demonstrated here in a set of homogeneous and heterogeneous phantoms with up to 5-fold acceleration factors.

## Summary of Main Findings/Short Synopsis

Magnitude contrast MR Elastography was developed on the basis of a motion-sensitizing magnetization preparation to subsequently make use of any type of imaging sequence, like UTE or ZTE, to mechanically characterize tissues, otherwise inaccessible with standard MRE.

## Introduction

Standard MRE pulse sequences are essentially constrained by the additional gradients required to get sensitized to motion. It yields long echo times and long acquisition times. By inserting motion encoding gradients (MEG) within a longitudinal magnetization preparation pulse sequence, we fully decouple high-order motion encoding from zero-order spatial encoding such that the MR signal magnitude, and not the phase, eventually carries the motion information before the image data acquisition has even started. The latter can then be carried over, free of any of the above constrains, by any type of pulse sequence.

## Theory

In DENSE-MRE, [1] the two MEG lobes were split before the RF excitation pulse, in between a pair of +90° and -90° RF pulses, and between the RF excitation and the image data acquisition. Hence, *TE* and *TR* could be shortened but not down to the respective sub-millisecond and millisecond ranges promoted by UTE [2] and ZTE [3] pulse sequences for short signal lifetime tissues. In the line of Diffusion Preparation, [4] Motion-Sensitized Driven-Equilibrium, [5] or Motion-Sensitizing magnetization Preparation or MSPrep, [6] we implemented an original MSPrep pulse with (Figure 1):

1. One +90° RF pulse with a 0° phase to flip the total magnetic moment in the transverse plane;
2. One motion encoding gradient lobe applied along one chosen encoding direction;
3. One 180° refocusing RF pulse with a 90° phase;
4. A second motion encoding gradient lobe identical to the first one.
5. One -90° RF pulse applied with a phase β at the preparation echo time to flip back the motion-encoded magnetization onto the longitudinal axis.

For perfect spin refocusing and no other sources of spin dephasing than the motion-induced Δϕ, the resulting longitudinal magnetization $M_z$ will be:

$$M_z=M_{z0}e^{-\frac{\Delta T_{MSPrep}}{T_2}} \cos \left( \Delta\phi - \beta\right)$$

$$M_z = M_{z0}e^{-\frac{\Delta T_{MSPrep}}{T_2}} \cos(\Delta\varphi - \beta) \qquad \text{Eq. 1}$$

where $M_{z0}$ is the initial longitudinal magnetization and $\Delta T_{MSPrep}$ is the total preparation duration. With subsequent spoiling gradients over $\Delta T_{Spoil}$ before initiating the image acquisition with an $FA$ RF excitation pulse, the expected amplitude of the transverse magnetization $M_{xy}$ is then:

$$M_{xy}\left(\Delta\phi\right)=\left[ M_0+\left( M_{z0} e^{-\frac{\Delta T_{MSPrep}}{T_2}} \cos \left( \Delta\phi - \beta\right) – M_0\right) e^{-\frac{\Delta T_{Spoil}}{T_1}} \right] \sin \left( FA\right) \tag{Eq. 2}$$

$$M_{xy}(\Delta\varphi) = \left[M_0 + \left(M_{z0}e^{-\frac{\Delta T_{MSPrep}}{T_2}} \cos(\Delta\varphi - \beta) - M_0\right)e^{-\frac{\Delta T_{Spoil}}{T_1}}\right] \sin(FA) \qquad \text{Eq. 2}$$

If Δ$\varphi$ remains small, then:

$$\cos \left( \Delta\phi - \beta\right) \simeq \cos\left(\beta\right) +\sin\left(\beta\right)\Delta\phi $$

$$\cos(\Delta\varphi - \beta) \approx \cos(\beta) + \sin(\beta)\Delta\varphi \qquad \text{Eq. 3}$$

In this case, if $β$ is well chosen and if $\left|\frac{\Delta\varphi}{\beta}\right| \ll 1$, |Δϕ /β| the recorded prepared MR magnitude images carry similar motion information than standard MRE phase images.

## Methods

MSPrep-MRE was demonstrated on four mechanically homogeneous phantoms (HoP) and a mechanically heterogeneous phantom (HeP) with five ellipsoidal inclusions (CIRS, Arlington, VA, USA). Reference Young's moduli were measured by shear wave elastography (SWE) with a 15-4 ultrasonic probe on an Aixplorer® system (Supersonic Imagine, Aix-en-Provence, France). MSPrep-MRE was carried in a 8-channel head coil on a GE Signa PET/MR 3.0 T (GE Healthcare, Waukesha, WI, USA). Pressure waves were remotely generated at 80 Hz, for HoP,

150 Hz and 180 Hz for HeP before being guided to the surface of each phantom and synchronized to the MRE sequence. MSPrep was implemented with β = 45° along three orthogonal spatial directions and for eight mechanical phases. Each preparation was followed, for HoP, by 200 repetitions of a 3D UTE acquisition with *FOV* = (100 × 100 × 126) mm$^3$, voxel = (1.5 mm)$^3$, *TE* = 12 μs and *TR* = 2 ms, and *FA* = 3° and, for HeP, by 150 repetitions of a 3D UTE acquisition with *FOV* = (120 × 120 × 240) mm$^3$, voxel = (1.3 mm)$^3$, *TE* = 12 μs and *TR* = 2 ms, and *FA* = 5°. Young's modulus maps were reconstructed by inversion of the Helmholtz equation after directional filtering for HeP while assuming **Eq. 3**.

## Results

MSPre-MRE magnitude images, displacement field maps and Young's modulus maps are presented in Figure 3 for HoP and on Figure 4 for HeP. The mean and standard deviation values of the Young's modulus are reported for SWE and MSPrep-MRE in Table 1 for HoPs and in Table 2 for HeP background. Averaged peak values are provided for the inclusions in HeP (Table 1).

## Discussion

MSPrep-MRE fulfills the basic requirements of any MRE pulse sequence. In HoPs, MSPrep-MRE provides fair measurements of the Young's modulus value that characterizes each phantom. On HeP, the five inclusions are clearly depicted in the Young's modulus maps whereas they are not directly visible in the UTE magnitude image (Figure 4, top row).

MSPrep-MRE Young's moduli are always lower than with SWE. This is probably related to the much higher mechanical excitation frequency used in SWE. [7]

In all these experiments, MSPrep-MRE total acquisition times are significantly reduced compared to standard MRE. In HoPs, MSPrep-MRE took 192 s whereas standard MRE would last at least 900 s, say 4.7 times longer. In HeP, at 150 Hz and 180 Hz, the acquisition of the same amount of data would have been 3.2, and 2.8 times longer in standard MRE.

## Conclusion

MSPrep-MRE seems to be a very promising way mechanically characterize tissues with short lifetime while reducing total acquisition times. It was made possible by decoupling MR motion encoding elements from MR imaging ones. This independence is ensured though a Motion-Sensitizing Preparation pulse, or MSPrep, which encodes motion directly in the subsequently acquired magnitude MR image.

## Acknowledgements

MRE experiments were performed on the GE Signa PET/MR 3.0 T platform of CEA/SHFJ, Orsay, France.

# References

1. Robert B, Sinkus R, Gennisson J-L, Fink M. Application of DENSE-MR-elastography to the human heart. Magnetic Resonance in Medicine. 2009;62(5):1155–63.
2. Tyler DJ, Robson MD, Henkelman RM, Young IR, Bydder GM. Magnetic resonance imaging with ultrashort TE (UTE) PULSE sequences: Technical considerations. Journal of Magnetic Resonance Imaging. 2007 Feb 1;25(2):279–89.
3. Weiger M, Pruessmann KP. MRI with Zero Echo Time. In: eMagRes [Internet]. John Wiley & Sons, Ltd; 2007
4. Koktzoglou I, Li D. Diffusion-Prepared Segmented Steady-State Free Precession: Application to 3D Black-Blood Cardiovascular Magnetic Resonance of the Thoracic Aorta and Carotid Artery Walls. Journal of Cardiovascular Magnetic Resonance. 2007 Jan 1;9(1):33–42.
5. Wang J, Yarnykh VL, Hatsukami T, Chu B, Balu N, Yuan C. Improved suppression of plaque-mimicking artifacts in black-blood carotid atherosclerosis imaging using a multislice motion-sensitized driven-equilibrium (MSDE) turbo spin-echo (TSE) sequence. Magnetic Resonance in Medicine. 2007;58(5):973–81.
6. Nguyen TD, Rochefort L de, Spincemaille P, Cham MD, Weinsaft JW, Prince MR, et al. Effective motion-sensitizing magnetization preparation for black blood magnetic resonance imaging of the heart. Journal of Magnetic Resonance Imaging. 2008;28(5):1092–100.
7. Yue J (2017) Magnetic Resonance and Ultrasound Elastography: simulation, experimental cross-validation, and application to liver characterization Université Paris-Sud.

# Figures

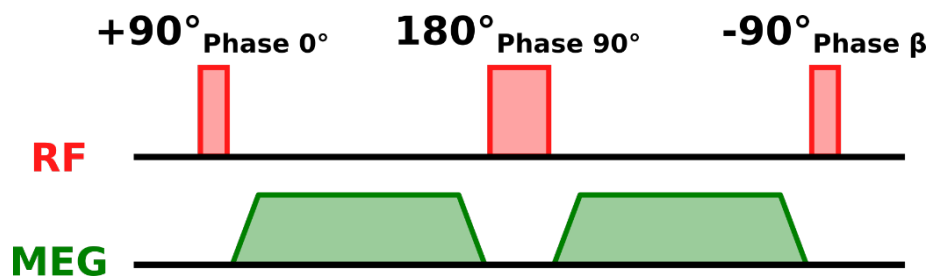

Figure 1: The Motion-Sensitizing Preparation (MSPrep) pulse.

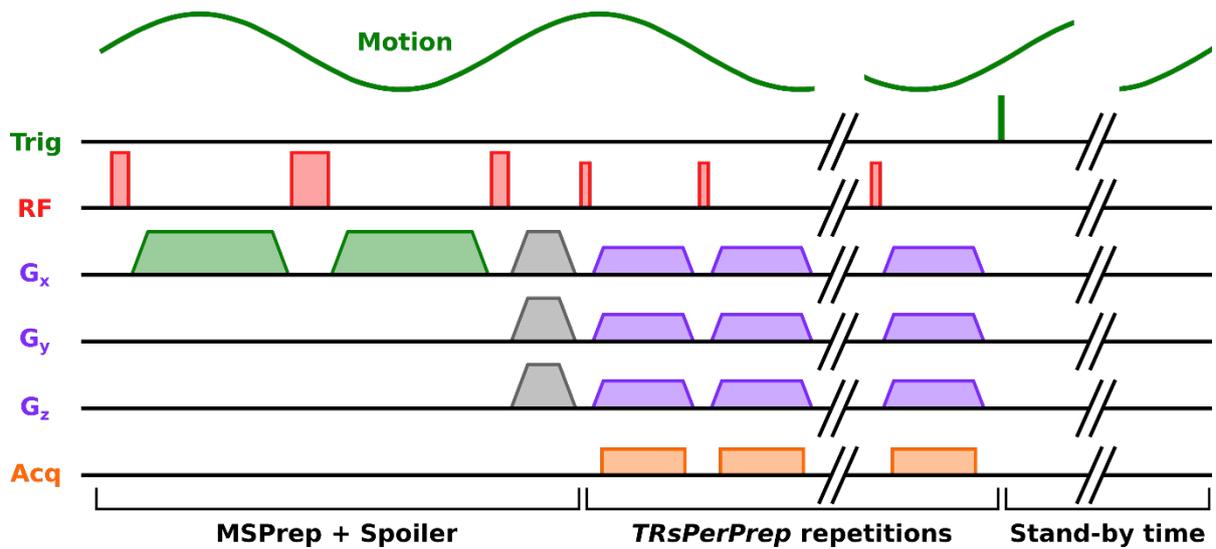

Figure 2: Representation of one segment of the UTE-based MSPrep-MRE pulse sequence and delimitation of its three main parts: the motion preparation and residual magnetization spoiling, the train of MRI repetitions and the stand-by time (200 ms and 150 ms for HoP and HeP acquisitions respectively).

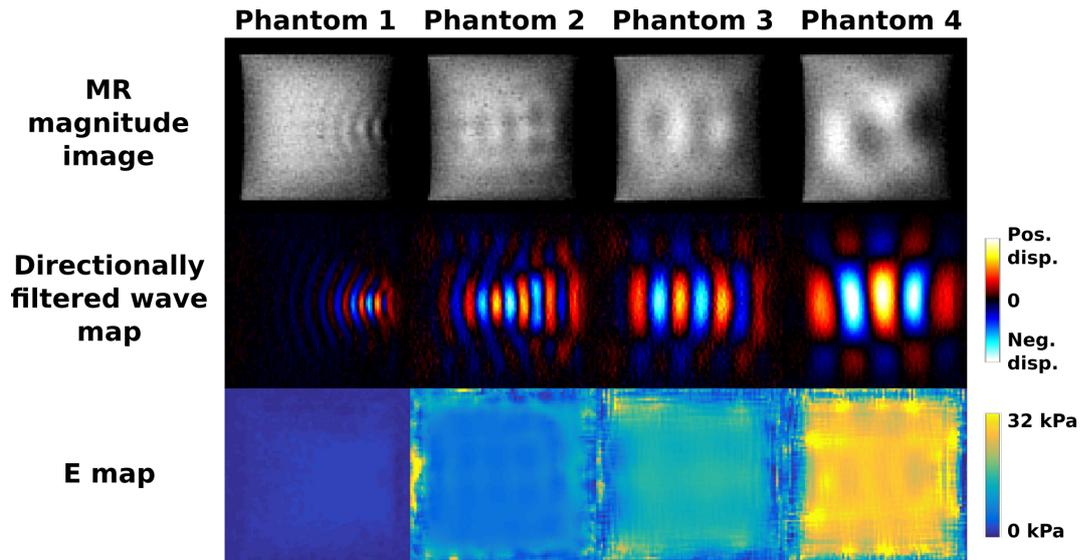

Figure 3: MR magnitude images, wave maps directionally filtered along the right-to-left direction and Young's modulus maps obtained for the four homogenous phantoms mechanically excited at 80 Hz.

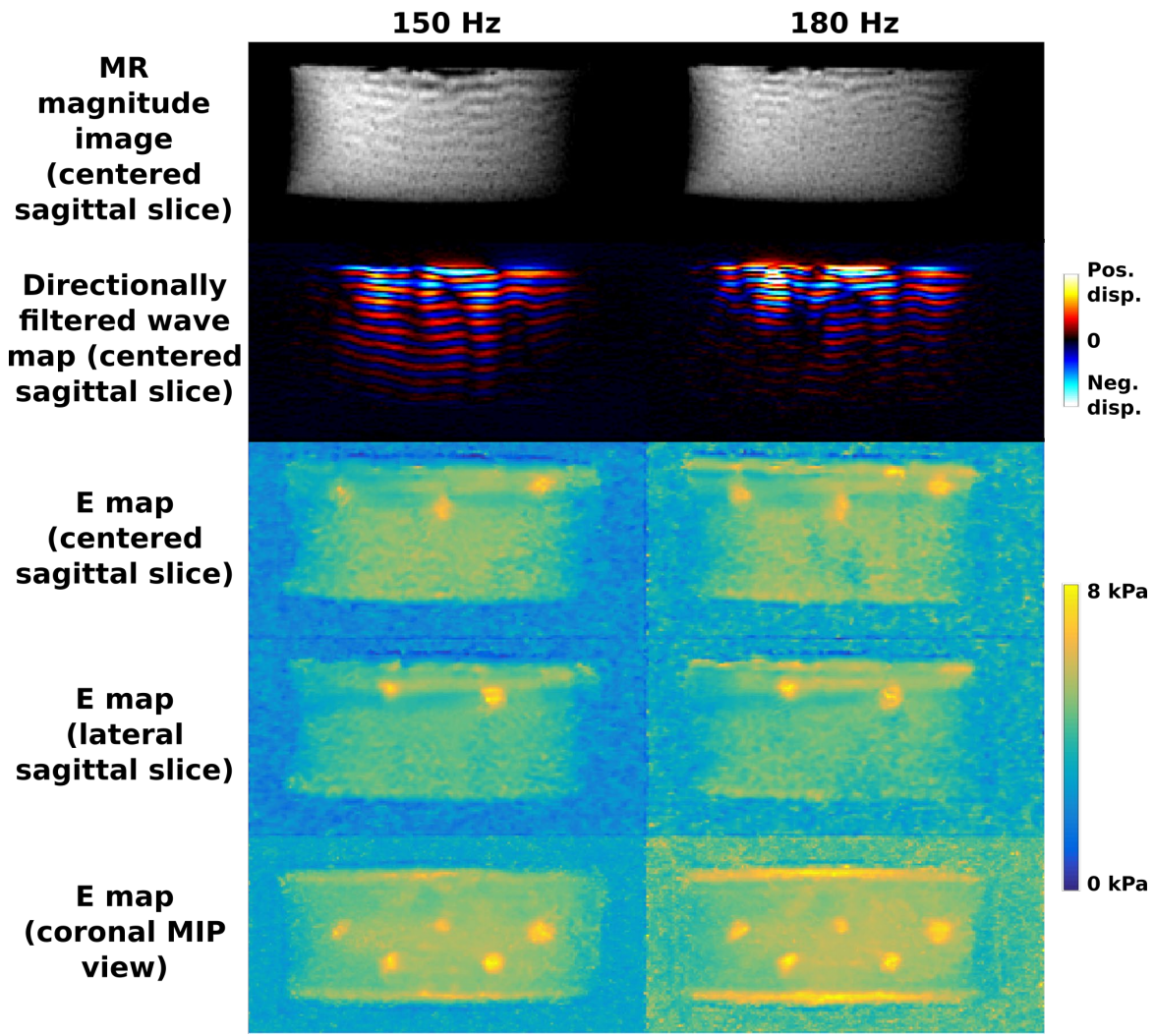

Figure 4: MR magnitude images, wave maps directionally filtered along the up-to-down direction and Young's modulus maps obtained for the heterogeneous phantom (HeP) at 150 and 180 Hz. Two sagittal planes with three and two inclusions are presented. A Maximum Intensity Projection (MIP) in the coronal view clearly depicts the five inclusions.

| Phantom | HoP 1 | HoP 2 | HoP 3 | HoP 4 |
|---|---|---|---|---|
| SWE-15-4 | (3.0 ± 0.5) kPa | (7.9 ± 0.6) kPa | (18.1 ± 0.7) kPa | (36.6 ± 2.6) kPa |
| MSPrep-MRE | (2.02 ± 0.3) kPa | (6.51 ± 0.7) kPa | (14.4 ± 0.8) kPa | (28.2 ± 1.8) kPa |

Table 1: Mean and standard deviation of the Young's modulus measured in four mechanically-homogenous phantoms (HoP) with shear wave elastography and MSPrep-MRE at 80 Hz.

| Heterogenous Phantom | Background | Inclusions |
|---|---|---|
| SWE-15-4 | (6.2 ± 0.2) kPa | (11.5 ± 0.5) kPa |
| MSPrep-MRE (150 Hz) | (5.01 ± 0.4) kPa | (8.03 ± 0.9) kPa |
| MSPrep-MRE (180 Hz) | (4.89 ± 0.5) kPa | (8.40 ± 0.6) kPa |

Table 2: Mean and standard deviation of the Young's modulus measured in the background and the five inclusions of a mechanically-homogenous phantoms (HeP) with shear wave elastography and MSPrep-MRE at 150 Hz and 180 Hz.